# Time-resolved predissociation of the vibrationless level of the B state of CH$_3$I

N. Thiré, R. Cireasa, D. Staedter, V. Blanchet*

[1]Université de Toulouse, UPS, 118 route de Narbonne, F-31062 Toulouse, France

[2]CNRS, Laboratoire Collisions Agrégats Réactivité, IRSAMC, F-31062 Toulouse, France

and

S.T. Pratt

Argonne National Laboratory, Argonne, IL 60439, US

Corresponding author :

V. Blanchet, fax # : 00 (33) 561 558 317, val@irsamc.ups-tlse.fr

**ABSTRACT**

The predissociation dynamics of the vibrationless level of the first Rydberg state 6s (B $^2$E) state of CH$_3$I has been studied by femtosecond-resolved velocity map imaging of both the CH$_3$ and I photofragments. The kinetic energy distributions of the two fragments have been recorded as a function of the pump-probe delay, and as a function of excitation within the umbrella and stretching vibrational modes of the CH$_3$ fragment. These observations are made by using (2+1) Resonant Enhanced MultiPhoton Ionization (REMPI) via the 3p$_z$ $^2$A$_2$" state of CH$_3$ to detect specific vibrational levels of CH$_3$. The vibrational branching fractions of the CH$_3$ are recovered by using the individual vibrationally state-selected CH$_3$ distributions to fit the kinetic energy distribution obtained by using nonresonant multiphoton ionization of either the I or CH$_3$ fragment. The angular distributions and rise times of the two fragments differ significantly. These observations can be rationalized through a consideration of the alignment of the CH$_3$ fragment and the effect of this alignment on its detection efficiency.

Two extra dissociation channels are detected: one associated with Rydberg states near 9.2 eV that were observed previously in photoelectron studies, and one associated with photodissociation of the parent cation around 15 eV.

# 1. Introduction

Time-domain studies of the predissociation of Rydberg states of methyl iodide have primarily been performed by monitoring the decay of the parent ion signal produced by photoionization of the Rydberg states. Using this approach, the vibrationless level of the first optically active Rydberg state, namely the B 6s[2] state, was found to have a picosecond lifetime,[1, 2] while higher vibrational levels of this state were found to have significantly shorter lifetimes.[3-5] Recently, we have studied the predissociation of the vibrationless level of the B 6s[2] state by time-resolved photoelectron spectroscopy,[6] and in the present paper, we extend this work by using fs-REMPI (Resonant Enhanced Multiphoton Ionization) to study the fragment branching fractions and translational energy distributions. The present results can be compared with the recent results of Gitzinger et al., which were obtained by using a very similar experimental set-up.[7]

The B 6s[2] state is derived from the promotion of $5p\pi$ electron into a 6s Rydberg orbital, leaving a ground state $CH_3I^+$ $^2E_{3/2}$ ion core. Predissociation of the 6s Rydberg state may occur by spin-orbit interaction with the continuum of the anti-bonding $\sigma^*$ lowest unoccupied molecular orbital (LUMO) localized on the C-I bond. This $\sigma^*$ orbital is responsible for the well-known A band. Two dissociation channels are energetically allowed, leading to a ground state $CH_3$ radical and an iodine atom in either the $^2P_{3/2}$ ground state or the $^2P_{1/2}$ spin-orbit excited state. Hereafter, these two states are labeled I and I*, respectively. The corresponding dissociation thresholds have been determined to be $2.366 \pm 0.013$ eV and $3.309 \pm 0.013$ eV by a state of the art experiment that combined a hexapole to select the initial state and a velocity map imaging spectrometer to measure the distribution of the fragments.[8] The energetics of the neutral and ionic states of the $CH_3I$ parent and its fragments are illustrated schematically in Figure 1, along with the excitation schemes used to prepare the $CH_3I$ and to detect its fragments. Vibrational excitation of the $CH_3$ is expected to populate the umbrella mode ($\nu_2$

~607 cm$^{-1}$) preferentially, owing to the transition from the nearly tetrahedral geometry of the CH$_3$ in CH$_3$I to the planar structure of the free CH$_3$ ground state. CH$_3$ fragments produced by photodissocation within the A band also show some activity in the $\nu_1$ symmetric stretching mode (~3004 cm$^{-1}$),[9, 10] and this may also occur following excitation to the B state.

Here we examine the predissociation of vibrationless level of the B-6s[2] state by using time-resolved photoion velocity map imaging. The CH$_3$ and I fragments were selectively ionized by REMPI schemes that allow the determination of the vibrational distribution of the CH$_3$ fragment. The main new results presented here are the time-dependencies of the CH$_3$ fragments as a function of the polarisation of the probe, the angular distribution of CH$_3$ and two new channels of dissociation detected on the iodine distribution. These two channels correspond to dissociation via Rydberg state at ~9 eV in a process that also produces a Rydberg fingerprint in the photoelectron spectrum, and to dissociation of the cation at an energy around 15 eV.

The remainder of the paper is organized as follows. A brief description of the experimental setup is given in Section 2, and the choice of probe wavelength is discussed in Section 3-1. Next, the time-resolved photoion signals integrated over kinetic energies and angle are presented in Section 3-2, and the CH$_3$ vibrational energy distribution is given in Section 3-3. The time dependences of the fragments as a function of their energies are summarized in Section 3.4. All the results are discussed in Section 4.

## 2. Experimental Details

The experimental setup is a standard combination of a continuous molecular beam coupled to a Velocity Map Imaging (VMI) spectrometer and has been described previously.[6] The laser is a 1kHz-3mJ/p Ti:Sapphire regenerative-amplified system (Amplitude System), delivering pulses with a central wavelength of ~805 nm and a Fourier-transform-limited full width at half-maximum (FWHM) duration of ~70 fs. The 1.8 mJ/pulse dedicated to femtochemistry is divided into two parts. One part is used to generate the pump beam by using crystals to produce the fourth harmonic at 201.2 nm with pulse energy of 1.5 µJ. This energy has been increased to 2.8 µJ for the investigation of the photodissociation of the cation. The second part is used to generate the probe beam by using a home-made Non-collinear Optical Parametric Oscillator (NOPA) with three stages of amplification. The NOPA delivers compressed pulses with 25 µJ of energy that are tunable between 510 and 700 nm. This output is then mixed with the 805 nm fundamental to generate tunable pulses between 313 and 340 nm with 7 µJ of energy, allowing the state-selective ionization of $CH_3$ with different numbers of quanta in the umbrella mode. The FWHM spectral bandwidths of 150 and 120 cm$^{-1}$ for the pump and probe, respectively, result in a correlation time around ~160 fs, somewhat shorter than the experimental FWHM of ~300+/-50 fs, depending on the probe wavelength. The pump and probe beams are focused onto the molecular beam by using 500 and 200 mm lenses, respectively. By measuring the spatial mode of the light pulses with a UV-CCD camera, the intensities of the two beams are found to be $< 6 \times 10^{11}$ W/cm$^2$. Iodine detection is accomplished simply by photoionizing with the second harmonic of the fundamental pulse, i.e., at 403 nm. The 403 nm light has a FWHM of 120 cm$^{-1}$, and a pulse energy of 35 µJ, corresponding to $3 \times 10^{12}$ W/cm$^2$. This energy has been reduced to 5µJ for the investigation of the photodissociation of the cation. In most of the experiments, the linear polarizations of the pump and probe are parallel to each other, and are also parallel to the face of the detector.

The continuous molecular beam is composed of a 10% $CH_3I$/Ar mixture, obtained by flowing 350 Torr of Ar through a $CH_3I$ sample held at -30° Celsius. This mixture expands through a 250 µm nozzle, and the resulting molecular beam is collimated by two skimmers. Under these expansion conditions, there is no evidence for the formation of $CH_3I$ clusters or $I_2$ in the mass spectrum or as additional features in the ion images, indicating that dimer formation is negligible.[5, 11-13]

In the interaction region, the molecular beam is crossed at right angles by the laser beams. The charged species created in this region are extracted perpendicularly to the plane defined by the laser and the molecular beams, through a 40 cm time-of-flight tube and are projected onto the detector, which consists of two micro channels plates (MCP) of 40 mm coupled to a phosphor screen. Transient ion signals are recorded by integrating the signal collected from the phosphor, which was subsequently preamplified and sent to a gated boxcar integrator. At each pump-probe delay, the signal is averaged over $3 \times 10^3$ laser shots. The raw images, recorded by a 12 bit digital video camera, are reconstructed by using Abel transformations to provide the 3-D distributions. At selected pump-probe delays, higher quality images were recorded by averaging over $2 \times 10^5$ laser pulses. The $CH_3$ images have been recorded with 4 kV applied to the repeller, which typically results in a resolution of ~100 meV at a fragment translational energy of 2 eV. The iodine fragments were also recorded at a repeller voltage of 4 kV to allow a direct comparison to $CH_3$ distribution. In addition, to improve the energy resolution, images of iodine were also recorded at a repeller voltage of 2 kV (~20 meV at 0.5 eV).

## 3. Results

The absorption spectrum of $CH_3I$ is dominated by a single sharp peak at 6.16 eV (201.2 nm), and there is no evidence for a absorption continuum within the laser bandwidth (150 cm$^{-1}$).[14] This situation ensures that only the v=0 level of B-6s[2] state is excited by one photon of the pump. The probe wavelengths are chosen to photoionize selectively the methyl fragments as a function of their internal energy, while the atomic I fragments are ionized nonresonantly by using 403 nm photons. Before presenting the results, we discuss the energetics of the ionization schemes at these different probe wavelengths.

### 3.1 Choice of the probe wavelength

Our optical set-up does not produce pulses of sufficient intensity at 304 nm to allow the state-selective ionization of I*, in contrast to the experiments of Gitzinger et al.[7] Instead, the iodine fragments are photoionized nonresonantly by using four photons at 403 nm. Several near resonances at three-photon energy could enhance this ionization process. These three-photon resonances include $^3P_1 6d[1]_{1/2} \leftarrow$ I* $^2P_{1/2}$ at 74425 cm$^{-1}$, $^1D_2 5d[2]_{1/2} \leftarrow$ I* $^2P_{1/2}$ at 74632 cm$^{-1}$, and $^3P_1 5d[1]_{3/2} \leftarrow$ I $^2P_{3/2}$ at 74587 cm$^{-1}$.

The CH$_3$ fragment is produced in the 2p$_z$ $^2A_2''$ electronic ground state and is probed by (2+1) REMPI via the 3p$_z$ $^2A_2''$ Rydberg state. The wavelength was tuned into resonance with the Q branch band heads, corresponding to absorption of two-photons at 333.45 nm for the $0_0^0$ band, at 329.5 nm for the $2_1^1$ band ($\nu_2$ is the umbrella mode), at 326.1 nm for the $2_2^2$ band, and at 322.8 nm for the $2_3^3$ band.[15, 16] The effective FWHM bandwidth for the two-photon absorption is 170 cm$^{-1}$, which is significantly smaller than the offset between successive sequence bands, thus allowing vibrational state selective ionization. This probe bandwidth is also sufficiently broad to overlap the entire Q branch within each umbrella band, and its duration is short enough to compete efficiently with the predissociation of the intermediate 3p$_z$ $^2A_2''$ Rydberg state into CH$_2$ + H. Note, however, that the bandwidth of the probe at the wavelength for the $0_0^0$ band does overlap the $1_1^1$ symmetric stretch band and the $3_1^1$ degenerate asymmetric stretch band, both of which occur at 2x333.9 nm.[17]

### 3.2 Time-resolved integrated photoion signals.

Figure 2 shows the femtosecond transients of the parent and fragment peaks observed in the time-of-flight mass spectrum recorded by using the current collected from the phosphor screen. Each ion signal is thus integrated over the translational energy and angular distributions. Figure 2a shows that the decay times of the CH$_3$I$^+$ signal recorded with the probe centered at 333.5 nm or 403 nm are the same even though a different number of probe photons is required for ionisation at these two wavelengths. This decay time corresponds to 1.31 ± 0.07 ps. These measurements are consistent with the 1.38 ± 0.2 ps lifetime measured in one-photon transitions;[1, 3] the 1.57 ps lifetime measured by using two photon excitation at 400 nm and multiphoton excitation at 800 nm;[2, 18] and the 1.5 ± 0.1 ps lifetime obtained by photoionization of the vibrationless level with 304 nm photons.[7]

These transients observed for the $I^+$ and $CH_3^+$ masses have a similar rise time to the parent ion signal, but decay to a non-zero plateau at longer times. The rapid rise and fast fall off are due to dissociative ionization to either $CH_3^+ + I$ or $CH_3 + I^+$, with energetic thresholds of 12.248 ± 0.003 eV [19] and 12.82 ± 0.02 eV,[20] respectively. The slower rise producing the plateau at longer times can be assigned to the photoionization of neutral fragments produced by the predissociation of the B-6s[2] state. This assignment is discussed in more detail below, and confirmed by using the velocity distributions determined in the imaging experiments. The transients can be fitted by considering two contributions, one for the dissociative ionization process, and a second rising component for the predissociation as given by:

$$S_{frag^+}(t) = a_0 + \left(1 + erf\left(\phi(t, T_e)\right)\right) \times b_0 e^{-\frac{t-t_0}{T_e}} + \left(1 + erf\left(\phi(t, T_r)\right)\right) \times c_0 \left(1 - e^{-\frac{t-t_0}{T_r}}\right)$$

Eq. 1

with the error function, $\phi(\Delta t, T)$, defined as

$$\phi(t,T) = \frac{2\sqrt{\ln 2}}{T_{cc}}\left(t - \frac{T_{cc}^2}{8\ln 2 \times T}\right).$$

Eq. 2

Here $T_e$ is the decay time of the parent and $T_r$ is the rise time of the fragment. The cross-correlation time $T_{cc}$ fixed to the measured value of 300 ± 50 fs and the $t_0$ a shift of appearance relative to the "zero delay" are determined by a non-resonant pump-probe ionization of NO. In Figure 2, the drastic assumption $T_r=T_e$ is made for most of the fits except for that from figure 2d. In this later, two different fits are done : one with $T_r=T_e$ (fit in blue) and one in which $T_e$ corresponds to the parent decay ($T_e$=1.31 ps) and $T_r$ corresponds to the rising time measured via the fs-REMPI velocity map imaging (see Table II, $T_r$=2.24 ps, fit in green).

**3.3 Energy distribution of photofragments at 8 ps**

Figure 3 shows velocity-map ion images for I and $CH_3$ recorded with a pump-probe delay of 8 ps. The single-color background has been subtracted from these images. Figure 3 also shows reconstructed slices of the three-dimensional velocity distribution obtained by using the Abel transformation. With an energy of excitation at 6.16 eV,[21] the maximum available energy is 3.80 eV for the I channel, and 2.86 eV for the I* channel. The transition from the ground state to the B-6s[2] state has a perpendicular charactere. Thus, for predissociation that is slightly faster than the rotational period of the parent, the fragments issued from a non-vibrating parent are expected to be preferentially ejected perpendicular to the polarization axis of the pump beam, as it can be seen for the image recorded for I fragments shown in Figure 3a.

In contrast, independent of the probe wavelength, the $CH_3$ images exhibit different angular distributions (Figures 3b and 3c). These distributions are far more anisotropic than the iodine image with minimum intensity at the poles and equators. This observation is at first surprising because the conservation of momentum in the dissociation process requires the angular distribution of the $CH_3$ fragments to mirror that of the I fragment. Indeed the predissociation at 201.2nm leads to I* fragment with J = ½ as we will discuss further. This fragment cannot be aligned; the I* detection efficiency should then be independent of angle, and the observed $I^+$ angular distribution should reflect the true angular distribution. The different observed angular distribution for the $CH_3^+$ image thus indicates the presence of an angular dependence of the detection efficiency of the $CH_3$. The lack of intensity at the poles of the $CH_3$ distribution suggests a lower detection probability for the fragments flying perpendicularly to the probe laser polarisation axis due to an unfavourable alignment of the $CH_3$ transition dipole moment and rotational angular momentum relative to the probe polarization. This alignment can be confirmed by recording an image with the probe polarization perpendicular to the pump polarization. An image recorded in this configuration is shown in Figure 3d, in which a strong $CH_3^+$ signal is observed perpendicular to the pump polarization. No such angular distributions were mentioned in the recent work of Gitzinger et al.[7] Alignment of the $CH_3$ fragment rotational angular momentum has been observed previously for photodissociation of $CH_3I$ within the A band,[22,23] and similar angular distributions were also attributed to alignment effects in the two-photon dissociation of Rydberg states in $CH_3Br$.[24] To our knowledge, the observations reported here represent the first detection of angular momentum alignment using broad band fs lasers and imaging techniques. These effects will be discussed in more detail in a forthcoming paper.

Figure 4 shows the total translational energy distributions extracted from the images for nonresonant ionization of I and $CH_3$ (4a and 4b, respectively), and for resonant ionization of $CH_3$ $v_2 = 0$, 1, and 2 (4e, 4d, and 4c, respectively), at a pump-probe delay of 8 ps. Both our previous work[6] and the work of others[7] indicate that the dominant dissociation channel for the B 6s[2] state is to $CH_3$ + I*. The images for state-selective ionization of $CH_3$ allow us to assign the dominant peak in the corresponding translational energy distributions and thus internally calibrate Figures 4 c-e in a straightforward fashion. An accurate value for that dissociation threshold ($D_0$=3.309 ± 0.013 eV) has been determined from studies of the A band.[8] With this assignment, the calibration of the nonresonant $CH_3$ and I translational energy distributions follows directly. While the absolute accuracy of the internal calibration is not expected to be as good as that obtained by using an external standard, for the present study the most important aspect is the relative calibration. This relative calibration is expected to be quite good, as all of the images were recorded under the same conditions. With this calibration, it is also clear from Figure 4 that there is no obvious contribution from dissociation to the lower energy $CH_3 + I(^2P_{3/2})$ products. This observation is consistent with our earlier results obtained from photoelectron spectra.[6]

The comparison of the translational energy distributions from the nonresonant I and $CH_3$ images is revealing. In particular, the I distribution in Figure 4a extends to significantly higher energy than the $CH_3$ distribution in Figure 4b, exceeding the thermodynamical threshold. Although Gitzinger et al.[7] plot their I and $CH_3$ distributions as a function of the translational energy of the individual fragments rather than the total translational energy of both fragments, transforming their I distribution to a total translational energy axis reveals a behavior analogous to that of our data. The present $CH_3$ distributions recorded with probe wavelengths of 400 nm and 800 nm are similar, indicating that these two detection schemes are essentially nonresonant. Interestingly, the translational energy distributions of iodine are different if the angular integration of the Abel image is limited to angles close to directions parallel or perpendicular to the pump polarisation (see Figure 4a). Because the I* fragment cannot be aligned, this difference indicates that two processes with different angular distributions are contributing to the I image and its translational energy distribution. Figure 5 shows the angular distributions obtained from $I^+$ images recorded with a pump-probe delay of 8 ps and corresponding to different translational energies (Figure 4a). The corresponding anisotropy parameters are $\beta = -0.549 \pm 0.005$ for total translational energy of iodine fragment between 2.2 eV and 3.0 eV, and $\beta = +0.89 +/- 0.02$ for fragment energies between 3 eV and 3.8 eV. While the process corresponding to lower translational energies clearly corresponds to the ionization of I* fragment produced by predissociation of the B-6s[2], the nature of the process producing higher translational energies is discussed in Section 4.2.

Figure 4c-e shows the energy distributions recorded using probe wavelengths of 326.1, 329.5 and 333.5 nm corresponding to the $2_2^2$, $2_1^1$ and $0_0^0$ bands, respectively, of the (2+1)REMPI transition involving the $3p_z$ $^2A_2''$ Rydberg state as a resonant step.[15, 25] Attempts to record an image with the probe wavelength tuned into resonance with the $2_3^3$ band at 322.8 nm were unsuccessful, suggesting that the predissociation process does not generate detectable amounts of CH$_3$($\nu_2$=3). This is confirmed by the CH$_3$ distribution recorded with a nonresonant probe. The typical widths of the translational energy distributions in Figures 4c-e vary between 165 and 85 meV. The vertical full lines in Figure 4 indicate the energy threshold expected for the fragment vibrational energy corresponding to bending excitation ($\nu_2$ progressions). The $\nu_1$ contribution is probed by (2+1) REMPI via $1_1^1$ band using 333.9 nm wavelength, which lies within the bandwidth of the 333.5 nm light used for the $0_0^0$ band. As already observed at 193.3 nm, in the predissociation of the $\nu_2$=2 level of the 6s[2] state,[26] the CH$_3$ distribution is characterized by significant excitation of the symmetric stretching mode $\nu_1$. In Figure 4, the dashed lines indicate the threshold for excitation to the $\nu_1+\nu_2$ combination mode, which was observed and assigned in previous fs studies of A and B band dissociations by Nalda et al.[10] and Gitzinger et al.[7] Nevertheless, transitions involving the $\nu_1+\nu_2$ bands are not spectroscopically known.

The vibrationally state-selected translational energy distributions of Figure 4c-e have been used to fit the translational energy distributions from both the I and CH$_3$ images (Figure 4a-b) recorded nonresonantly, and thus provide a determination of the CH$_3$ vibrational distribution. This distribution is summarized in Table I and Figure 6. The vibrational distribution found for CH$_3$ recorded off resonance at 403 nm or 800 nm is the same. In contrast to Gitzinger et al.[7] who directly used in their fits the translational energy distributions measured resonantly, as only one component, we have used normalised functions resulting from the fit of each vibrational component observed in Figure 4. This approach completely decouples the contributions of each vibration and thus eliminates the possible artifacts in the relative intensities produced by variations in the detection efficiencies of the $\nu_1$ and $\nu_2$ components. The $\nu_2 = 1$ feature is the largest component of both vibrational distributions presented in Figure 6, but there is also significant intensity in the $\nu_2 = 0$ component. In contrast, in the vibrational distribution reported by Gitzinger et al., the $\nu_2 = 0$ component is dominant. This difference might just result from the fitting procedure.

### 3.4 Rise time of the neutral fragments.

Figure 7 shows the time dependence of the fragment signals as a function of the pump-probe delay. The curves for I and CH$_3$ have been recorded in two and three independent scans, respectively. These time-dependences are fitted by the rising function of Eq.1 (i.e., the term that includes $c_0$), except for that from Fig. 7-b. The latter decay process is fitted by a cross-correlation function and a decaying exponential similar that multiplies the term in $b_0$ of Eq.1. The time constants and time shifts extracted by fitting the data are summarized in Table II.

*Iodine fragment*

The iodine translational energy distribution is characterized by three different contributions: one corresponding to fragments emitted perpendicular to the pump polarisation at 2.8 eV and two corresponding to fragments emitted parallel to the pump polarisation at 3 eV and 6.4 eV. These three components have different time-dependences, as shown in Figure 7a, 7c and 7b respectively.

*Perpendicular component around 2.8 eV – Figure 7a*

The iodine fragments emitted between 80°-100° (perpendicular to the pump polarisation) appear within 1.17+/-0.06 ps and have a translational energy of about 2.8 eV (Figure 7a). As expected by taking into account the shift of appearance, these fragments have an overall appearance time corresponding to the decay time of the parent ion (Figure 2a) or either the rising function of the Figure 2b.

*Parallel component around 3 eV – Figure 7c*

The time dependence of the iodine fragments emitted parallel to the pump polarisation within a 20° angle and with an energy around 3 eV (see Figure 7c) can be split in two parts: a step function with a rising edge corresponding to the laser cross-correlation and an increasing function with a rise time on the picosecond scale. Note that the perpendicular contribution with β=-0.549 will lead to a contribution in the 20° parallel distribution of about 36% of what is emitted perpendicular but centered on 2.75 eV. Figure 4a recorded at 8 ps shows a percentage roughly 27% at 2.75 eV to compare to the expected 36%, while at 3 eV, around 2.5 more iodine atoms are produced parallel instead of perpendicular. This observation leads us to conclude that another process besides the predissociation from the v=0 of B-6s[2] state, takes place on a picosecond time scale, and that this process is not observed in the $CH_3$ product channel. To enhance this contribution, image were recorded with the pump energy increased to 2.8 µJ and the probe decreased to 5 µJ. Figure 8 shows then two Abel inverted images recorded at 1 ps and 8 ps as well as the energy distribution obtained by integration over an 20° angle around the pump polarisation and compared to the energy distribution emitted perpendicularly to the pump polarisation at 8 ps.

*Parallel component around 6.4 eV –Figure 7band Figure 9*

The energy distribution of iodine fragments emitted along the pump polarisation at high energy is shown in Figure 9. The time-dependence of this iodine shown in Figure 7b can be fitted by a cross-correlation function with an exponential decay on time scale 1.65+/-0.05 ps slightly longer that the $CH_3I^+$ decay 1.31 ps. We assigned this component to iodine fragment produced around 9.24 eV by the absorption of one probe photon from the v=0 level of B 6s[2] state. This (1+1') excitation path was already identified in the photoelectron spectrum via a Rydberg fingerprint onto the 6p[2,4] and 7s[2,4] states.[6] The vibrational energies in these Rydberg states are 1.92 eV in 6p [2] and 1.21 eV in 7s[2], and these states are expected to have very short lifetimes. Indeed the lifetimes of their vibrationless levels have been measured to be shorter than 150 fs.[27] These two ultrafast radiationless transitions would produce fragments with total available energies of 5.93 eV for the I* channel and 6.88 eV for the I channel, respectively. The production of iodine in its ground state can come from the interaction with the repulsive $^1Q(E)$ as predicted by the calculations for energy larger than 6.8 eV.[28] The anisotropy parameter obtained by fitting the angular dependence of the iodine translational distribution between 4.5 and 7 eV at a delay of 700 fs is β= 1.89+/-0.05. The kinetic energy of the $CH_3$ fragments associated with this process is so high that the corresponding ring in the image is larger than the diameter of the detector, and is thus not observed. To a first approximation, the lifetime of this component should reflect the lifetime of the v=0 of B 6s[2] state, since this state is an intermediate stepping stone in accessing the 9.4 eV resonances. It is noteworthy that this process occurs within the probe pulse duration and requires five photons of the probe with two resonant steps.

*Methyl fragment*

The methyl produced in $(\nu_1, \nu_2)$=(0,1) vibrational state and detected with a probe pulse centered at 329.5 nm has been recorded as a function of the pump-probe delay, as well as all the methyl fragments ionized near-resonance at 403 nm. The energy distribution does not change in time and results in the distribution shown in Figure 4d, which was fitted by two Gaussian distributions: one peaking at 2.7 eV and corresponding to one quantum excitation of the umbrella mode $(\nu_1, \nu_2)$=(0,1) (REMPI (2+1) transition at 329.5 nm) and the second one around 2.35 eV that we assign as Gitzinger et al.[7] to an excitation of the combination mode ($\nu_1$, $\nu_2$)=(1,1). As noted above, this would indicate that within the broad bandwidth of the probe pulse, a REMPI (2+1) transition probing this combination mode is expected in this wavelength region, however this has never been identified spectroscopically. Surprisingly both of these components have rise times that are considerably longer than any other timescales measured ($T_r$=2.2 ps, see Table II), especially when taking into account the shift of appearance time. The increase in the rise time is not an experimental artefact, as the same time dependence was obtained for three different experimental runs. To confirm this rise time, the methyl fragments ionized with a non-resonant probe pulse at 403 nm were also collected as function of the delay. These data also show a longer rise time ($T_r$=2 ps) than the iodine and the parent ion as seen in Figure 10.

## 4. Discussion.

In general, our data are in good agreement with the recently published results of Gitzinger et al. The observation that only I* fragments are detected from the predissociation of v=0 of the B 6s[2] state confirms the conclusions of the previous works.[7,26,29,30] This conclusion can be understood easily for the B-6s[2] vibrationless level by considering the crossing of the potential surfaces of the B-6s[2] state and the 4E($^3A_1$) and $^2A_2$ ($^3A_1$) valence states along the C-I stretching coordinate. Both the 4E and $^2A_2$ valence potential surfaces cross the B 6s[2] surface near its equilibrium geometry, and both diabatically correlate with the $CH_3$+I* dissociation limit.[28,31] The predissociation is caused by the spin-orbit coupling. The present vibrational state distribution differs slightly from the one of Gitzinger et al., but can be understood in terms of the different fitting procedures used for the $CH_3$ translational energy distributions.

## 4.1 The Time dependence of CH$_3$ fragments

We have measured for the first time, the temporal dependence of CH$_3$ fragment produced in the dissociation of the v=0 level of B-6s[2] state. The rise time for the signal corresponding to CH$_3$ ($\nu_1$, $\nu_2$)=(0,1) umbrella mode is 2.24 +/-0.12 ps, which is considerably longer than expected based on that of the corresponding I* signal. In order to rationalize this observation we propose a model which accounts for the CH$_3$ detection sensitivity as a function of the relative alignments of transition dipoles used to photodissociate the CH$_3$I and to photoionize the CH$_3$ fragment with respect to the polarizations of the pump and the probe. In the following discussion, it is assumed that the C$_{3v}$ symmetry group is preserved during the predissociation of CH$_3$I. The transition dipole moment of the B 6s[2] ← $^1$A$_1$' transition at 201.2 nm is perpendicular to the C$_3$ axis of the CH$_3$I and thus preferentially selects molecules with the C$_3$ axis lying in the plane perpendicular to the pump polarisation and parallel to the time-of-flight. Thus, if the dissociation is prompt, the C$_3$ axis of CH$_3$ fragment is (at least initially) parallel to the C$_3$ axis of CH$_3$I, and perpendicular to the polarization axis of the probe light, which is parallel to the polarization of the pump in this configuration. However, the transition dipole moment of the two-photon 3p$_z$ $^2$A$_2$" ←2p$_z$ $^2$A$_2$" step in the (2+1) REMPI used to detect the CH$_3$ is parallel to the C$_3$ axis of the CH$_3$. Consequently, for a prompt dissociation, the initial ionization probability is expected to be almost zero because the probe polarisation is perpendicular to the dipole moment of detection transition. However, the present dissociation is not a prompt process, but rather a predissociation taking place over 1.3 ps. This process is thought to involve a slow crossing onto the σ* surface, followed by rapid (<100 fs) dissociation on that surface. Therefore, one factor that can diminish this C$_3$ axis alignment effect is the rotation of the parent molecule before the dissociation takes place. Following this, the C$_3$ axis of the CH$_3$I will be no longer perpendicular to the probe polarisation, and the CH$_3$ fragments will be ejected over a greater range of angles relative to the probe polarization. Our main hypothesis is that the resulting position of the $C_3$ axis relative to the polarization of the probe laser has the potential to modify the appearance of this fragment. We will now discuss this in more detail.

Although the sample is cooled in a molecular beam, the rotational temperature can not be ignored. We will label $T_{rot}*$ the temperature corresponding to the angular energy causing $CH_3I$ to leave the plane selected by the pump transition and this corresponds to rotation about a $C_2$ axis. This rotation is the one that contributes most to the loss of anisotropy in the detection process. This $T_{rot}*$ can be deduced from the angular distribution of the I* fragment. If it is assumed that the B 6s[2] ← $^1A_1'$ transition is purely perpendicular, we assume that the deviation of the value of β from the limiting value of -1 to -0.549+/-0.005 (Figure 5) is due to rotation of the parent molecule outside of this plane selected by the pump transition. By treating classically the parent rotation, the angular distribution of the iodine fragment for a one-photon perpendicular transition can be written:[32, 33]

$$\Gamma(\theta) = \frac{1}{4\pi}\left(1 + \beta(T_e) P_2(\cos\theta)\right)$$
$$\text{with } \beta(T_e) = \beta(0) \frac{1+(\omega T_e)^2}{1+4(\omega T_e)^2} \text{ and } \beta(0) = -1 \quad (3)$$

For which the rotational energy is $E_{rot} = \frac{I_b \omega^2}{2}$ with the main moment of inertia for the C-I rotation $I_b = 111.8 \times 10^{-47}$ kg.m$^2$ and the predissociation time $T_e$ fixed to 1.3 ps in agreement with the decay time of the parent. Once integrated over the possible rotational energies taking into account their Boltzmann distributions, we get

$$\Gamma(\theta) \propto \frac{\sqrt{\pi}\left(3-\cos^2(\theta)\right)}{16 T_e \mu} + \frac{\pi\left[1-\phi(\mu/2)\right]\exp(\mu^2/4)}{8 T_e}\left(\sin^2(\theta) - \frac{\left(3-\cos^2(\theta)\right)}{4}\right) \quad (4)$$

With $\mu^2 = \frac{I_b}{2T_e^2 k_B T_{rot}^*}$ and $\phi(x)$ is the error function. This angular dependence can then be fitted by $\left[1+\frac{\beta}{2}(3\cos^2(\theta)-1)\right]$, as done for the experimental results. Figure 11 gives the anisotropy parameter β for the predissociation time 1.3 ps. The anisotropy parameter β=-0.549 results from $T_{rot}^* = 62$ K, to which we can associate an averaged rotational period around the C$_2$ axis $\left(\frac{2\pi\sqrt{\pi I_b}}{\sqrt{2k_b T_{rot}^*}}\right)$ of 9 ps. Note that the extracted rotational period is not particularly sensitive to the predissociation time, and the rotational period changes by only ~1 ps for an increase in predissociation time of 100 fs.

Assuming that the predissociation time is 1.3 ps, the C$_3$ axis will start to be aligned with the probe polarization after a quarter of a rotational period, i.e., ~2.25 ps. The observed rise time of the CH$_3$(ν$_2$=1) of 2.24 ps is consistent with this process. As far as we know the only angular dependence reported for the methyl fragments resulting from the Rydberg excitation is the study by Van Veen et al. with an excitation at 193 nm that populated the ν$_2$=2 of the B6s[2] state via $2_0^2$ transition.[30] The reported anisotropy parameter of the methyl fragment was measured to be around β = -0.36±0.05 while that for the iodine was around β = -0.72, as expected for a shorter lifetime of 760 fs. [3].

In order to insure a more efficient detection of the fragments flying perpendicularly and thus measure a less biased value of the predissociation time, one should rotate the probe polarization, such that to be perpendicular to that of the pump. This is illustrated in Figure 10 for a photoionization probe centered at 403 nm: the measured CH$_3$ rise time is 1.359 +/- 0.072 ps when the probe polarization is perpendicular to that of the pump, and 2.02+/-0.15 ps when the probe polarization is parallel to that of the pump. While the first value is similar to the measured decay time of the parent ion shown in Fig. 2a, the latter is similar to the value of $T_r$=2.24 +/-0.12 ps obtained when state selected CH$_3$ ($\nu_1,\nu_2$)=(0,1) was detected using pump and probe polarizations parallel to each other (Figure 7d). Indeed, although ionization at 403 nm is non-resonant, this four photon process is expected to have parallel character due to near resonance with the A$_1$'(5d) at 75000 cm$^{-1}$ (for the third photon).

The bandwidths measured for the translational distributions of fragments $\nu_2$=1 (160 meV) and $\nu_2$=2 (121 meV) in Figure 4 c-d, are significantly narrower than the translational distributions measured by Gitzinger et al.[7] This observation might be related to their slightly smaller value β = -0.5 for the anisotropy of the I* fragment. In particular, both observations are consistent with the conclusion that the expansion of the molecular beam (here a continuous beam and in Gitzinger work a pulsed beam) can affect the time-scale of detection of the methyl and the observed angular distribution.

### 4.2 The iodine cation emitted parallel around 3 eV.

The iodine distributions of Figure 4 show that some fragments are produced with translational energies that exceed the available energy for dissociation to CH$_3$ + I* following excitation at 201.2 nm. For example, the fragments with translational energy of 3.50 eV observed in Figure 4a exceed the available energy of 2.86 eV by approximately 22%. This unexpectedly high translational energy observed for the iodine fragment is not mirrored by the CH$_3$ co-fragment distribution. A similar behavior is observed in the data of Gitzinger et al., where an excess energy of 23% can be estimated from their Figure 2a.[7] We can decompose our iodine distribution integrated over 360° in two components : one emitted perpendicular to the pump polarisation (~61%) and another one emitted parallel (~30%-see Figure 4a). The perpendicular component is associated with the predissociation of the B 6s[2] state, and is consistent with the CH$_3$ energy distribution (see Figure 6). The surprisingly high kinetic energy fragments are associated with the parallel component as shown in Figure 8a. The time dependence of this parallel iodine is shown in Figure 7c, as well as in Figure 8b for different energies of pump and probe. We will examine now the possible assignment of this contribution.

If this fast I signal resulted from dissociation into $CH_3$ + I, rather than I*, following excitation at 201.2 nm, the available energy would be sufficient to explain the observation. However, the observed kinetic energy of the I would indicate that it is produced in conjunction with $CH_3$ ($v_2$ = 6,7), which would then be observed in the nonresonant $CH_3$ image (Figure 4b). The absence of this $CH_3$ as well as the observation that the angular distribution of this $I^+$ is peaked along the pump polarisation (Figure 8a), eliminates this potential explanation for the origin of fast I signal.

One way to produce iodine ions without methyl cations is via the dissociation of $CH_3I^+$ into $CH_3\left(X\,^2A_2''\right)+I^+(^3P_2)$ with a (0 K) threshold 12.82+/-0.02 eV above the $CH_3I$ ground state.[20] The energetics of the cation are summarized in Figure 1. Photoionization of $CH_3I$ by 2 photons at 201.2 nm is known to leave the cation mostly in the v=0 of the ground state $X\,^+E_{3/2}$ (at 9.538 eV).[6] Subsequent absorption of one probe photon will then increase the total excitation energy to 12.61 eV, just 355 meV above the threshold for dissociation to $CH_3^+$ + $I(^2P_{3/2})$. The first electronic excited state of $CH_3I^+$ is the $A(^2A_1)$ state whose minimum lies 2.4074 eV above the cation ground state. Transitions to the $A(^2A_1)$ state from the ground state are allowed with a transition dipole moment perpendicular to C-I bond.[34] This $A(^2A_1)$ state correlates diabatically to the $CH_3\left(X\,^2A_2''\right)+I^+(^3P_2)$ asymptote. However, for low vibrational levels, fast internal conversion takes place back to the ground state, leading to the dissociation of the cation into $CH_3^+$+I. Excitation by two extra probe photons will reach 15.7 eV (2+2'-$E_{kin}$(e-)). The second state of $CH_3I^+$ with $^2A_1$ symmetry lies around 15.22 eV, and is described as an excitation from the HOMO to the LUMO of the ion.[20] Another state occurs in this energy range, corresponding to the $B^2E_{3/2}$ state at about 15 eV, which is populated by a transition from a valence orbital into the partially filled HOMO.[20] The $B^2E_{3/2}$ state undergoes Jahn-Teller distortion, and the transition to the B state from the ground state of $CH_3I^+$ is broadened by internal conversion to the $A(^2A_1)$ state, followed by a dissociation into $I^+ + CH_3$.[35,36] The branching ratios measured in PEPICO experiment (photoelectron-photoion coincidence) at 15.7 eV are 40 % to $CH_3 + I^+$ and 60% to $CH_2I^+ + H$,[35] while the total energy released in the dissociation taking place at 15.8 eV is around 2.93 eV.[36] Photodissociation of $CH_3I^+$ by two photons has been already studied for wavelengths ranging from 427-720 nm.[34,37-40] These studies focused primarily on the low energy part of the A state by detecting $CH_3^+$ fragment. Only Goss have discussed the $I^+$ fragments appearance.[37] The two-photon transition $B(^2E_{3/2}) \leftarrow X(^2E_{3/2})$ can take place though a dipole transition $A_1 \times A_1$ or either $E \times E$. The first one will lead to $I^+$ emitted parallel to the probe polarisation while the second to $I^+$ emitted perpendicular to the probe polarisation which will thus be masked by the dominant contribution of the B-6s[2] state predissociation. Halogen atoms emitted parallel to the C-X axis following absorption into the B state has been already observed in the photodissociation of $CH_3Br^+$[41] and $CH_3Cl^+$[42] cations. Note as well that in Figure 5, the angular dependence of $I^+$ emitted parallel to the polarization axis is better fitted by assuming a two photon transition (blue line) than a one photon transition (black line): this fit supports our conclusion that the fast $I^+$ is produced by:

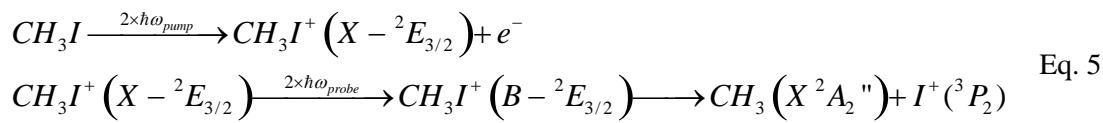

$$CH_3I \xrightarrow{2 \times \hbar\omega_{pump}} CH_3I^+\left(X - {}^2E_{3/2}\right) + e^-$$
$$CH_3I^+\left(X - {}^2E_{3/2}\right) \xrightarrow{2 \times \hbar\omega_{probe}} CH_3I^+\left(B - {}^2E_{3/2}\right) \longrightarrow CH_3\left(X\,{}^2A_2"\right) + I^+({}^3P_2)$$

Eq. 5

Because the resonant step in the 1 + 1 ionization of $CH_3I$ is a perpendicular transition, the C-I axis of the $CH_3I^+$ will initially be aligned perpendicular to the probe axis, and thus the two-photon parallel transition to the B state will not be efficiently excited until the rotation of the ion has brought the C-I axis parallel to the probe polarization. As in the case of the $CH_3$ detection discussed above, this rotation will take a while as observed on Figure 7(c). The maximum translational energy expected for this channel is 2.9 eV, while the energy observed at long delay for iodine emitted parallel to the probe is 3 eV (see Figure 8). This energy difference can be explained if some of the parent cations are produced with internal vibrational energy in the photoionization by two pump photons that are produced with some vibrational energy in the two pump photoionization. This contribution is clearly visible on Figure 8a for which the pump energy has been increased to enhance the production of cation by the pump alone, and the probe energy has been reduced to reduce the detection of the neutral iodine fragments produced by the predissociation from the neutral B state (around 3eV).

## IV. CONCLUSION

We have investigated the predissociation of $CH_3I$ from the vibrationless level of the B[2] 6s Rydberg state by imaging as a function of time the angular and translational energy distributions of the fragments. No I fragment has been detected leading us to conclude that the quantum yield of I* is close to 1. The dipole moment of this transition is perpendicular to the C-I axis leading to the ejection of iodine atoms mostly perpendicularly to the pump polarization. Instead, when the pump and probe polarization are parallel to each other and to the detector plane, the $CH_3$ image angular distribution exhibits four-fold symmetry due a low detection sensitivity of the fragments flying perpendicularly to the pump laser polarization. This observation together with the fact that the angular distribution changes when changing the probe polarization indicates an alignment of $CH_3$ rotational angular momentum. Using resonant multiphoton ionization for the $CH_3$ fragments we could determine the vibrational energy distribution. The $CH_3$ products have excitation primarily in the umbrella mode as expected and observed in most of the dissociations of methyl halide. This distribution is peaked at $\nu_2=1$, with some minor activity in the $\nu_1$ symmetric stretching as well. For this pump and probe polarization parallel configuration, while the iodine appears on a timescale in agreement with the decay time of the parent, the rise time of the main CH$_3$ fragment ($\nu_2$=1) is almost twice longer. We interpret this difference in the rising times of the two fragments as a result of a different sensitivity in the detection step. By considering the relative alignments of the transition dipoles used to photodissociate CH$_3$I and to photoionize the CH$_3$ fragment with respect to the polarizations of the pump and the probe, we propose a model in which the rotation of parent during predissociation alters the initial unfavorable alignment of the fragment C$_3$ axis to enable a delayed detection as observed experimentally. Both CH$_3$ angular distribution and the increased rising time are reflection of alignment effects and will be discussed in a forthcoming paper. By investigating their changes as a function of different relative configurations of the pump and probe polarizations, we will be able to to shed more light onto this unexpected observations.

Two other contributions are observed in the I* kinetic energy distribution, corresponding to fragments emitted along the pump polarization. They appear on different timescales and are produced by multiphoton processes. One corresponds to the decay of 6p and 7s Rydberg states populated by a (1+1') excitation and it was detected on the photoelectron spectrum as well.[6] Both I and I* seem to be produced by dissociation of one of these highly vibrationally excited Rydberg states. The present measurement gives an upper limit for the dissociation lifetime of one of these states, as being around 100 fs, namely the duration of the probe pulse. The second contribution comes from the dissociation by two probe photons of the CH$_3$I$^+$ produced with two photons of the pump onto CH$_3$+I$^+$. As with the detection of the methyl fragment resonantly, this dissociation channel is also affected by the rotational temperature of the parent molecule.

**Acknowledgments**

This work was supported financially by the ANR COCOMOUV, the ANR HARMODYN and L'Université Paul Sabatier via three different BQR. S.T.P. thanks the CNRS for supporting invited research position in the LCAR. S.T.P. was also was supported by the U.S. Department of Energy, Office of Science, Office of Basic Energy Sciences, Division of Chemical Sciences, Geosciences, and Biosciences under contract No. DE-AC02-06CH11357. R.C. gratefully acknowledges the European Union for the award of an Intra-European Marie Curie fellowship through the contract MOLCOTUV-041732. D.S. acknowledges the European Union for its PhD fellowship from the ITN-ICONIC. We thank Elsa Baynard, Laurent Polizzi and Stéphane Faure for their expert technical assistance and Dr. Lionel Poisson for lending us his image analysis software.

**TABLE I** Vibrational distribution at 8 ps resulting from the fit performed on the iodine and CH$_3$ translational energy distributions recorded with a probe centered 403 nm or 800 nm. The translational energy distribution of the iodine fragments that is fitted is the one obtained by integrating over the angles within 20° of the perpendicular to the pump polarization.

| % | $(\nu_1,\nu_2)=(0,0)$ | $(\nu_1,\nu_2)=(0,1)$ | $(\nu_1,\nu_2)=(0,2)$ | $(\nu_1,\nu_2)=(1,0)$ | $(\nu_1,\nu_2)=(1,1)$ |
|---|---|---|---|---|---|
| I$_{perp}$ | 30±3 | 30±3 | 15±2 | 10±2 | 4±2 |
| CH$_3$ at 403nm | 36±1 | 44±1 | 9±1 | 7±1 | - |
| CH$_3$ at 800nm | 28±2 | 56±2 | 4±2 | 10±2 | 2±1 |
| Ref {Gitzinger, 2010 #2539} | 37 | 28 | 24 | 7 | 4 |

**TABLE II** Time dependences extracted from the fits of the energy resolved signals of Figure 7 or of Figure 9. The fits have been performed with a cross-correlation time around $300 \pm 50$ fs. $I_{perp}$ and $I_{para}$ correspond to the intensity of iodine collected perpendicularly and parallel to the pump polarization within a 20° angle. In Figure 7, the pump and probe polarizations are parallel, while in Figure 9, they can be parallel or perpendicular.

|  | Time constant $T_r$ /ps | Shift of the appearance $t_0$/fs |
|---|---|---|
| Figure 7 - Pump-probe para. | | |
| (a) I perp ~2.8 eV | 1.17+/-0.06 | 193+/-20 |
| (b) I para ~6.4 eV | Decay:1.65+/-0.05 | 164 +/-10 |
| (f) I over 360° | 1.05+/-0.05 | 83+/-19 |
| (d) $CH_3$ $v_2$=1 | 2.24 +/-0.12 | 307+/52 |
| (e) $CH_3$ $v_2+v_1$ ? | 1.57+/- 0.11 | 910 +/- 60 |
| Figure 9 | | |
| $CH_3$ @403 nm Pump-probe para. | 2.02+/-0.15 | 157+/-65 |
| $CH_3$ @403 nm Pump-probe perp. | 1.36 +/- 0.07 | 249+/-36 |

**FIGURE CAPTIONS**

Figure 1. Schematic diagram of the electronic states, dissociation limits, and photoionization thresholds encountered after excitation of the vrationaless level of the B state at 201.2 nm. The nascent $CH_3$ fragments are probed by (2+1) REMPI (green arrows) with typical wavelength around 330 nm and iodine fragment is probed by four-photon ionization at 403 nm (red arrows).

Figure 2. Time-resolved transients recorded after excitation of the origin band of the B state at 201.2 nm: (a) the parent $CH_3I^+$ with the probe centered at 403 nm (red triangles) and 329.5 nm (blue crosses). The cross-correlation time is determined from the signal for the photoionization of NO. (b) $I^+$ produced by ionization at 403 nm. $CH_3^+$ produced by ionization (c) at 333.5 nm that allows REMPI via $0_0^0$ and (d) at 329.5 nm via $2_1^1$. See the text for more details on the fits.

Figure 3. Abel transformations at 8 ps delay of (a) iodine fragments recorded with a 403 nm probe with the probe polarization parallel to that of the pump. The repeller voltage is fixed at 2 kV. Raw images of $CH_3$ collected with a repeller voltage at 4 kV recorded at 8 ps with a probe centered at (b) 333.5 nm and (c) 403 nm, both with the probe polarization parallel to that of the pump. Fragment of $CH_3$ recorded with a 403 nm probe with (d) probe perpendicular to the pump polarization. The images have all been symmetrized and the backgrounds have been subtracted.

Figure 4. Normalized energy distributions plotted as a function of the translational energy of the center of mass and recorded for different probe wavelengths at 8 ps pump-probe delay.

(a) Full square: Iodine fully integrated over 360° of the Abel image. Circle: Iodine integrated over a 20° range around the perpendicular to the pump polarisation with a weight of 0.61. Triangle: Iodine integrated over a 20° range around the direction parallel to the pump polarisation with a weight of 0.3. The 0.61 and 0.3 weights have been obtained by fitting the total distribution (over 360°). The full line corresponds to the sum of the perpendicular and parallel contributions, and should be compared to the full integration over 360°. The vertical line at 2.86 eV corresponds to the threshold limit at one photon for $CH_3+I^*$.

(b) $CH_3$ ionized by four photons at 403 nm off resonance (empty circle) or seven photons at 800 nm (filled square).

(c-e) $CH_3$ detected by (2+1) ionization (c) at 326.1 nm via the $2_2^2$ resonance of the $3p_z\,^2A_2''$ state, (d) at 329.5 nm via the $2_1^1$ resonance of the $3p_z\,^2A_2''$ state, (e) at 333.5 nm via the $0_0^0$ resonance of the $3p_z\,^2A_2''$ state. The Gaussian fits in red and blue are used in the fit of the vibrational distributions shown in Figure 6. The full vertical lines correspond to the translational energy expected for $(n \times \nu_2)$ while the dashed ones correspond to $(n \times \nu_2 + \nu_1)$ progressions, respectively.

Figure 5. Angular distribution of iodine at 8 ps pump-probe delay and recorded at the translational energy corresponding to the maximum of signal between 2.2-3 eV (square, β=-0.549+/-0.005), and at higher energy between 3 to 3.8 eV (circle, black line with β=+0.89+/-0.02 or blue line $\beta_2$=+1.02+/-0.01 and $\beta_4$=-0.33+/-0.01). The angle 180° corresponds to iodine emitted along the pump polarization.

Figure 6. Fits of the vibration distribution recorded at 8 ps delay using (a) iodine integrated over 20° perpendicular to the pump polarization (Figure 4a) and (b) CH$_3$ recorded off-resonance at 403 nm (Figure 4b). The vibrational distributions are summarized in Table I.

Figure 7. Time dependences of the fragments.

(a) Iodine fragments integrated over 20° around the perpendicular to the pump polarization and for translational energies between 2.3 and 3.3 eV. These data have been recorded in two scans.

(b-c) Iodine fragments recorded over 20° around the direction parallel to the pump polarization and for translational energies within a 2 eV window centered (b) at 6.4 eV and (c) at 3 eV

(d) CH$_3$ fragments recorded with a probe centered at 329.5 nm. The y-scale corresponds to the intensity of the component $\nu_2$=1 recorded via the REMPI $2_1^1$ and extracted from a fit of the energy distribution performed for each delay. These data have been recorded in three scans.

(e) CH$_3$ fragments recorded with a probe centered at 329.5 nm. The y-scale corresponds to the intensity of the component assigned energetically to $\nu_2+\nu_1$ and extracted from a fit of the energy distribution done for each delay.

(f) Iodine fragments integrated over 360° and within a 2 eV window centered at 3 eV.

Figure 8. Iodine fragments recorded with a higher pump energy at 2.8µJ and a lower probe energy at 5µJ to enhance the dissociation dynamics taking place in the ion continuum. The pump and probe polarizations are parallel to each other and perpendicular to the time-of-flight axis. (a) Abel images recorded a 3kV for the repeller voltage for two different delays. Clearly, at 1 ps the outer ring is aligned along the pump and probe polarization while at longer delay a new component appear with roughly the same energy but pointing perpendicular to the pump polarization. This new component corresponds to the fragments that have been photoionized after the predissociation from the 6s-(B $^2$E) Rydberg state and hides the parallel contribution. (b) The translational energy distribution recorded for iodine fragments ejected within a 20° angle of the pump polarization as a function of the pump-probe delay and compared to the one recorded at 10 ps within a 20° angle perpendicular to the pump polarization (black plot).

Figure 9. Translational energy of the iodine fragment obtained by integration of the Abel transformation over 20° around the pump polarization and for different pump-probe delays. The integration over the component centered at 6.4 eV leads to the time-dependence shown in Figure 7b.

Figure 10. Time dependency of the $CH_3$ fragments ionized at 403 nm and produced with a energy distribution between 2.3 and 3.2 eV (see Figure 4b). The pump polarization is perpendicular to the TOF axis for both data, while two different polarizations of the probe have been tested, namely parallel (black squares) or perpendicular (red circles) to the pump. For comparison, the time dependency of $CH_3$ detected on resonance at $(v_1,v_2)=(0,1)$ shown already in Figure 7d is as well reported (blue triangles). The parameters deduced from the fit are listed in Table II.

Figure 11. The left axis shows the anisotropy parameter β as a function of $T_{rot}*$ of $CH_3I$ for a predissociation time of 1.3 ps. The horizontal line at β=-0.549 corresponds to the anisotropy

measured in Figure 5 for iodine emitted with $E_T$=2.2-3 eV. The right axis shows the rotational period of CH$_3$I as a function of $T_{rot}$*. For $T_{rot}$*=62K, the period is 9 ps.

**Figure 1**

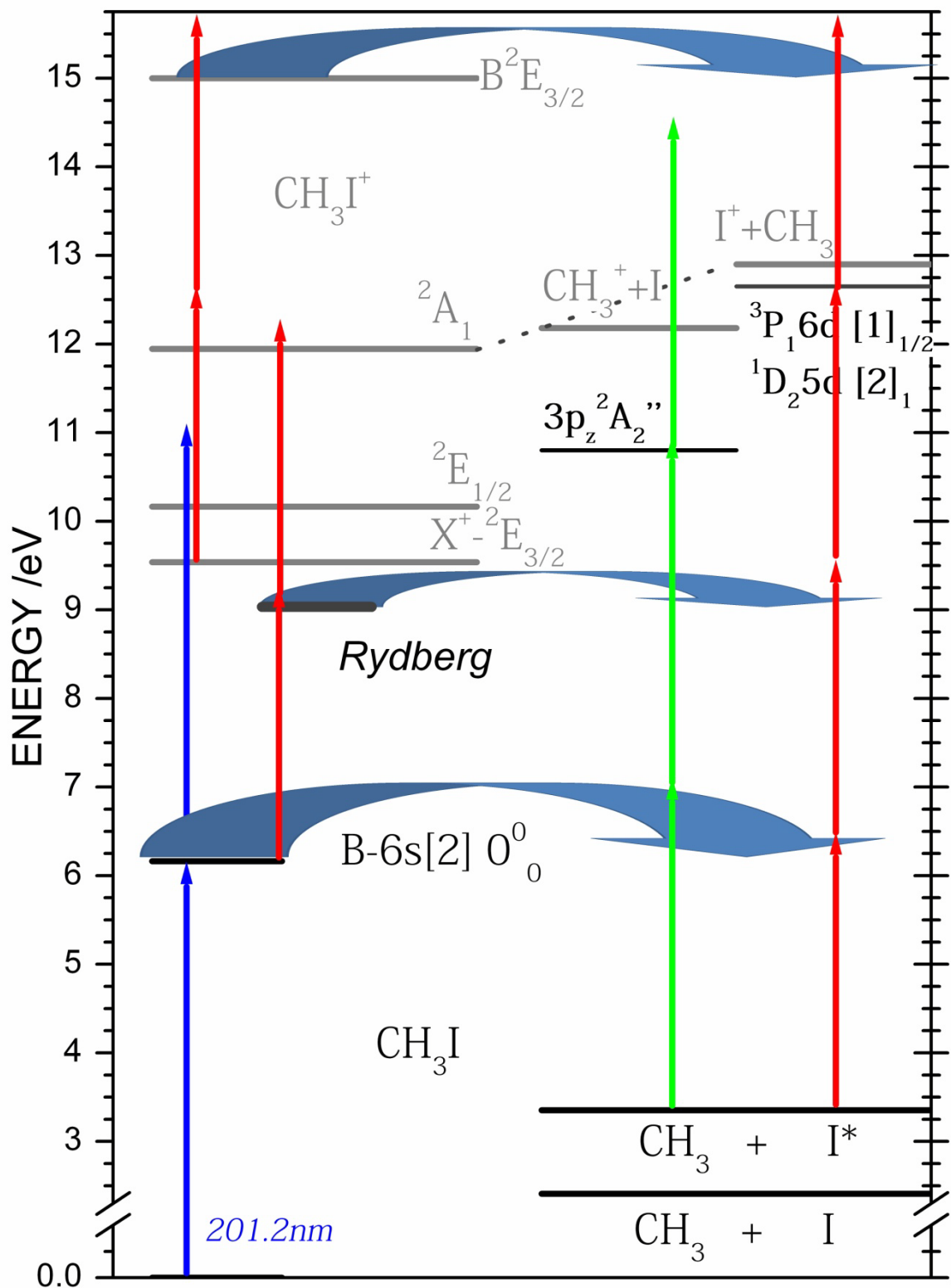

**Figure 2**

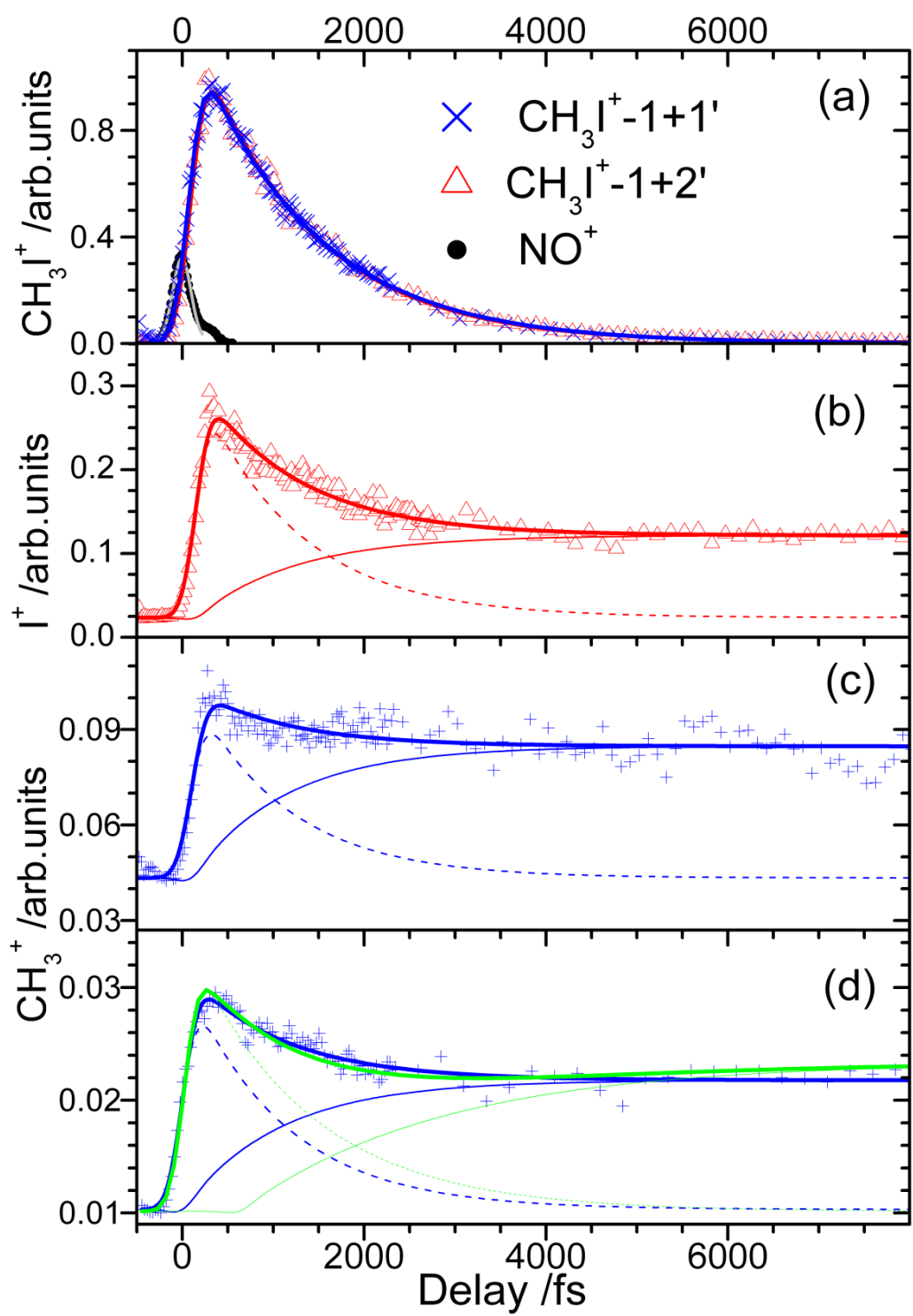

**Figure 3**

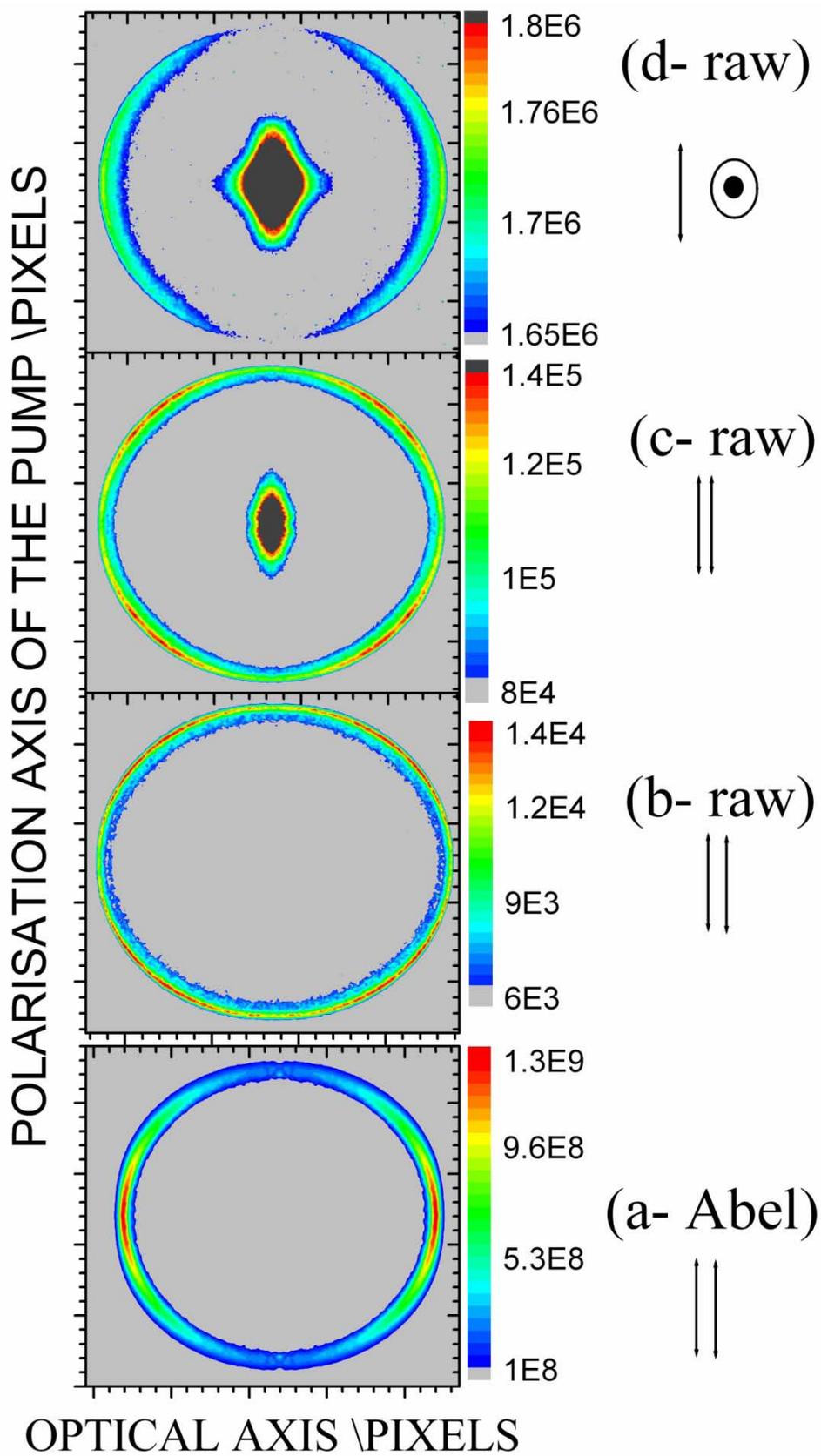

**Figure 4**

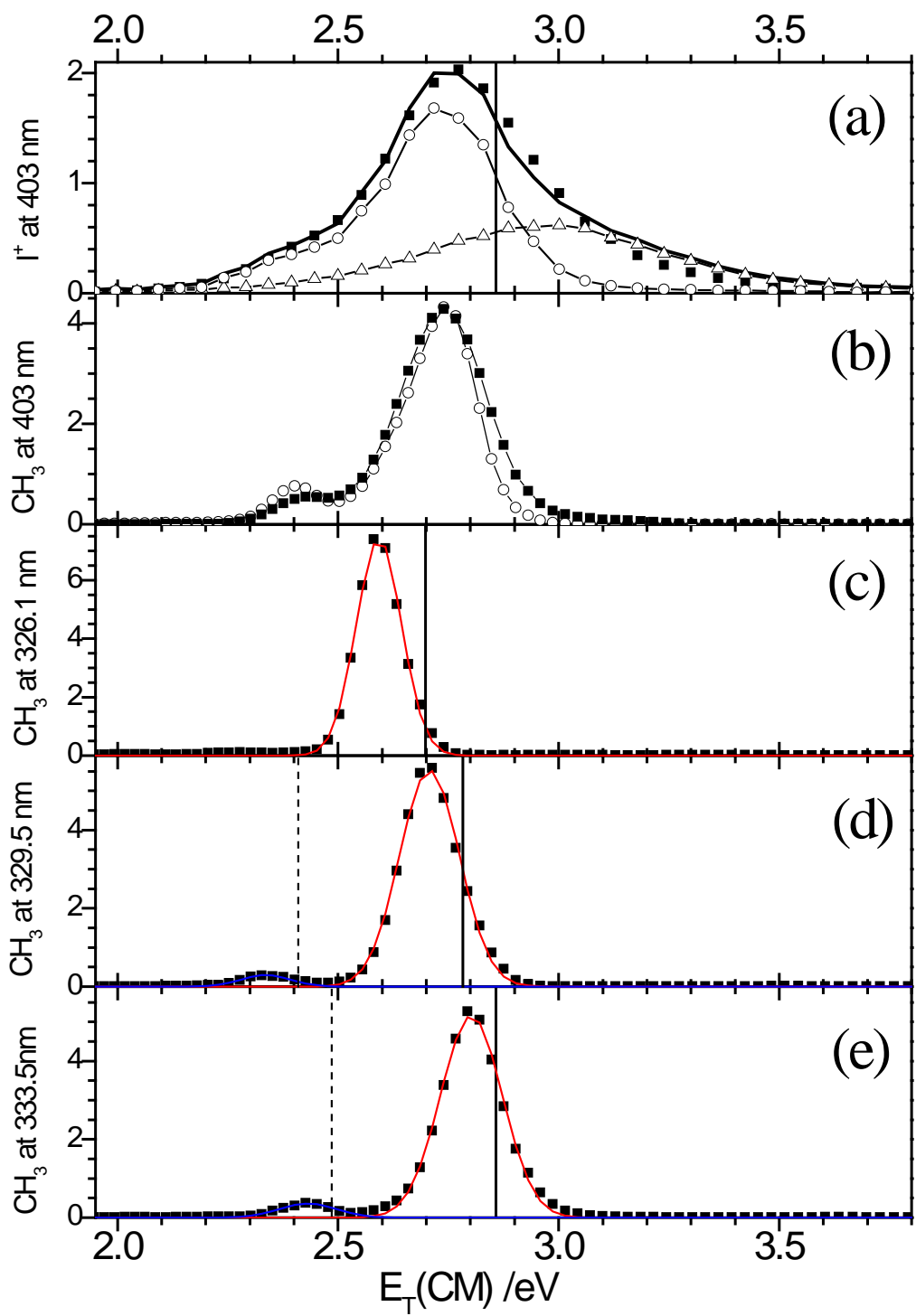

**Figure 5**

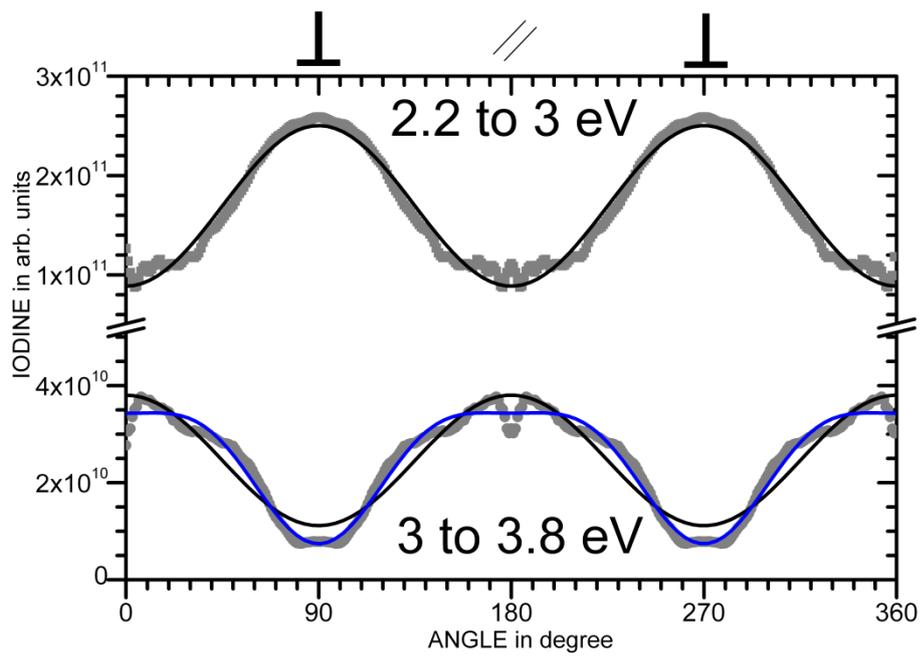

**Figure 6**

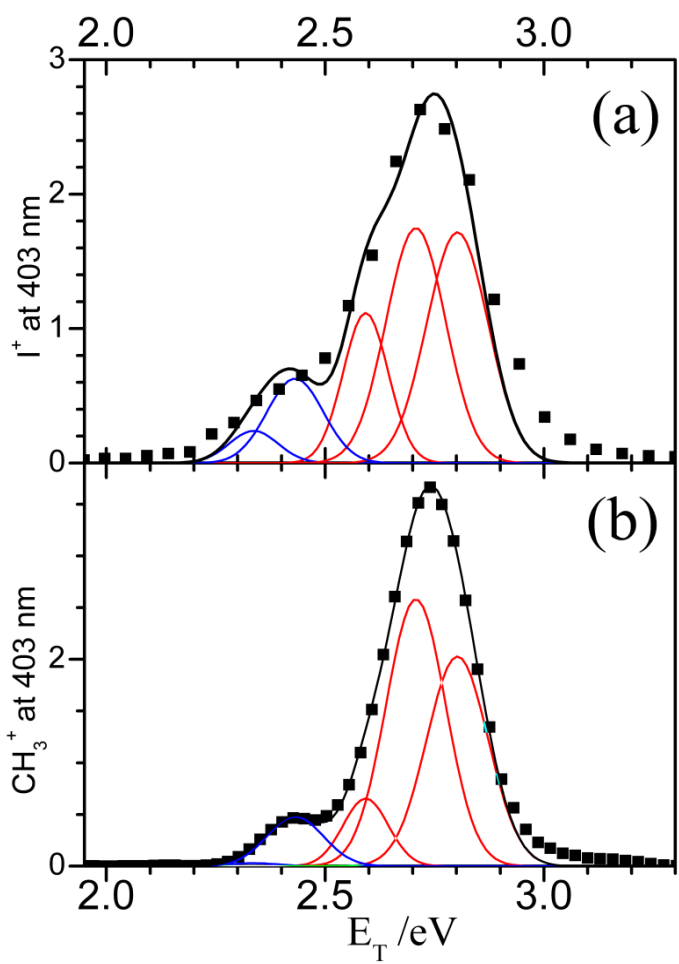

**Figure 7**

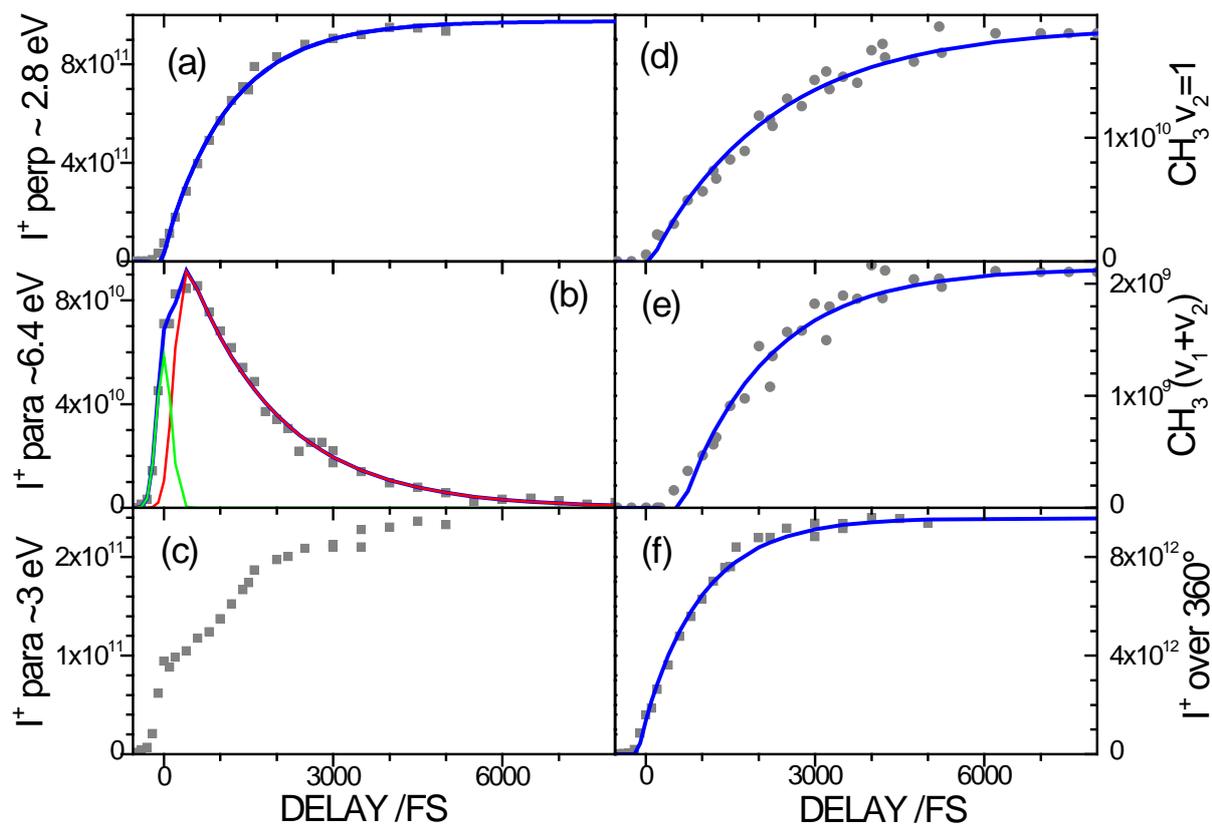

**Figure 8**

(a)

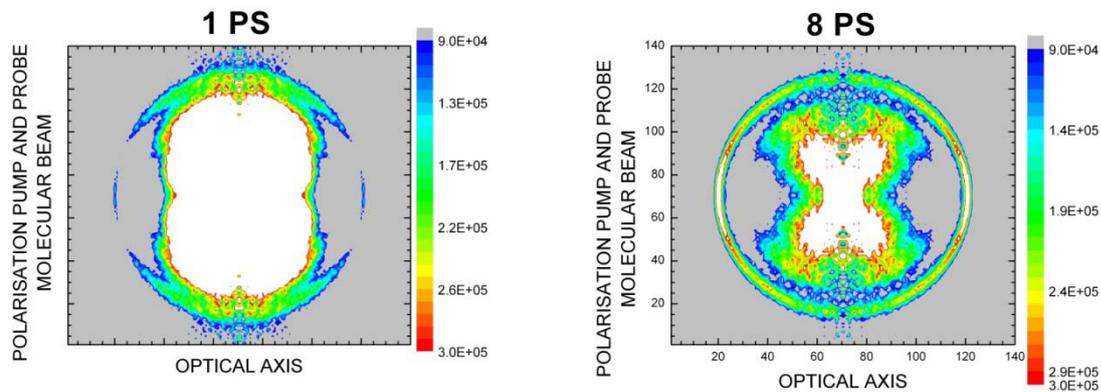

(b)

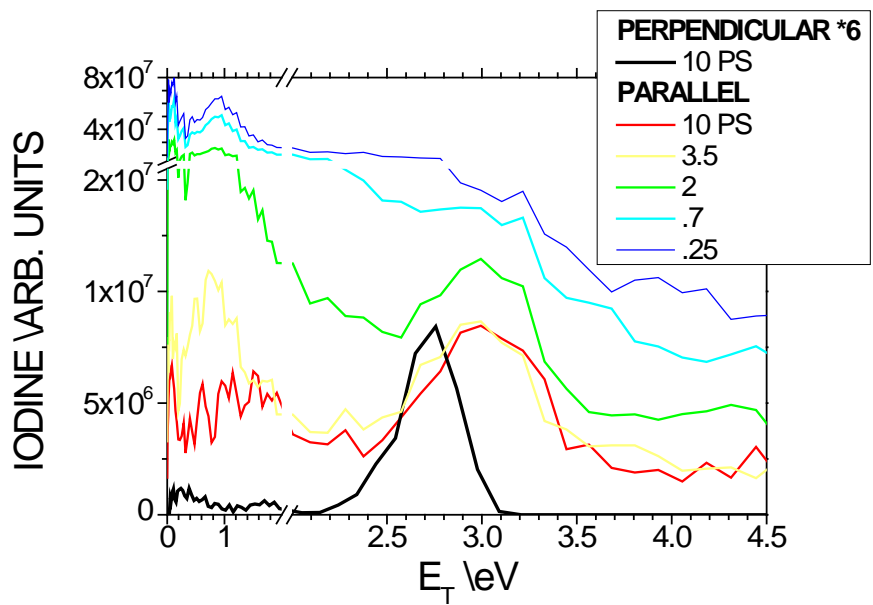

**Figure 9**

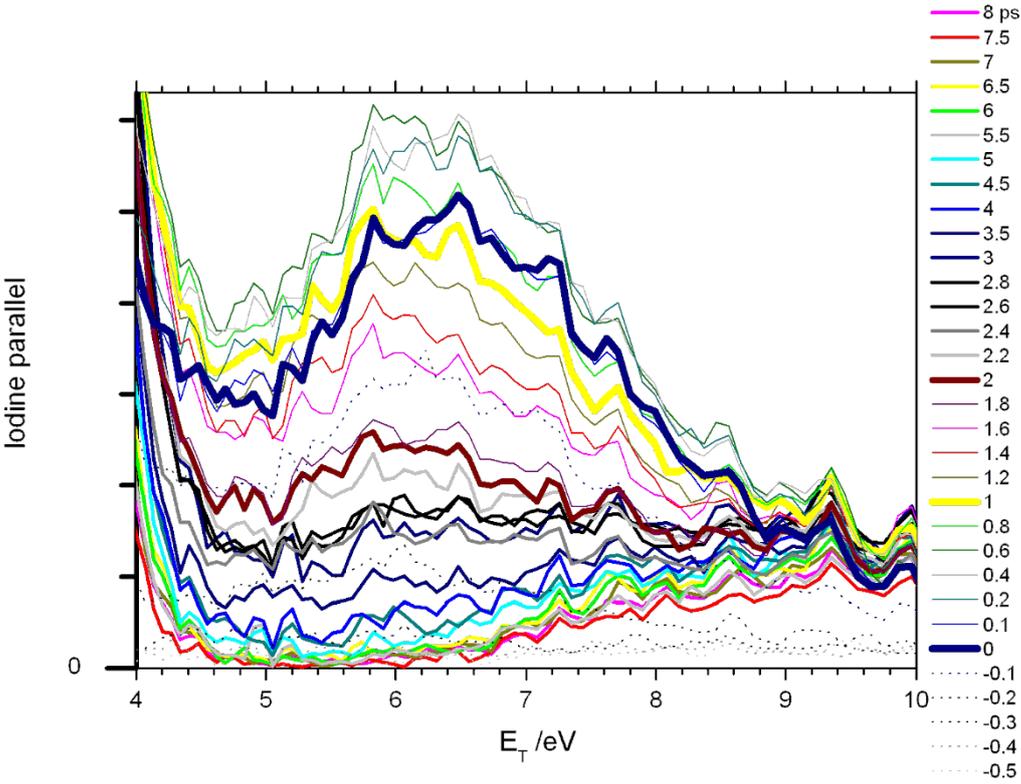

**Figure 10**

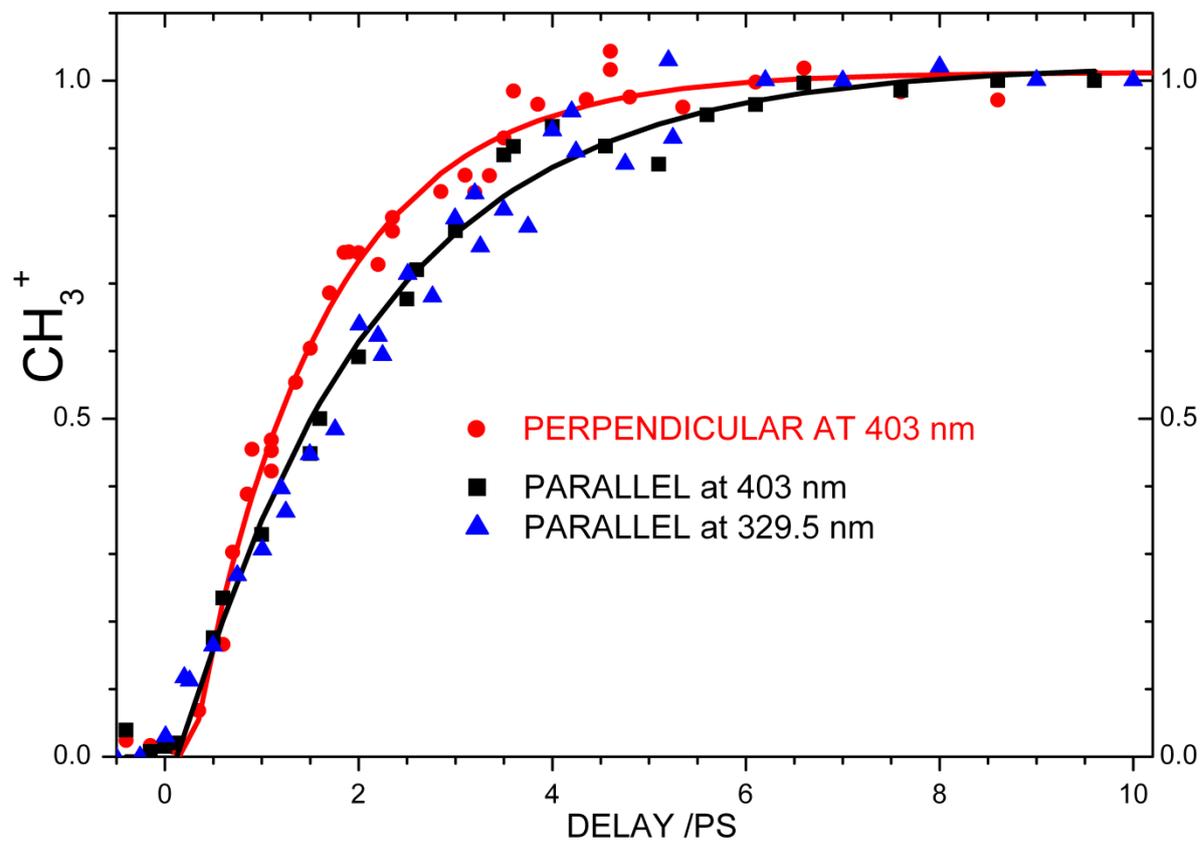

**Figure 11**

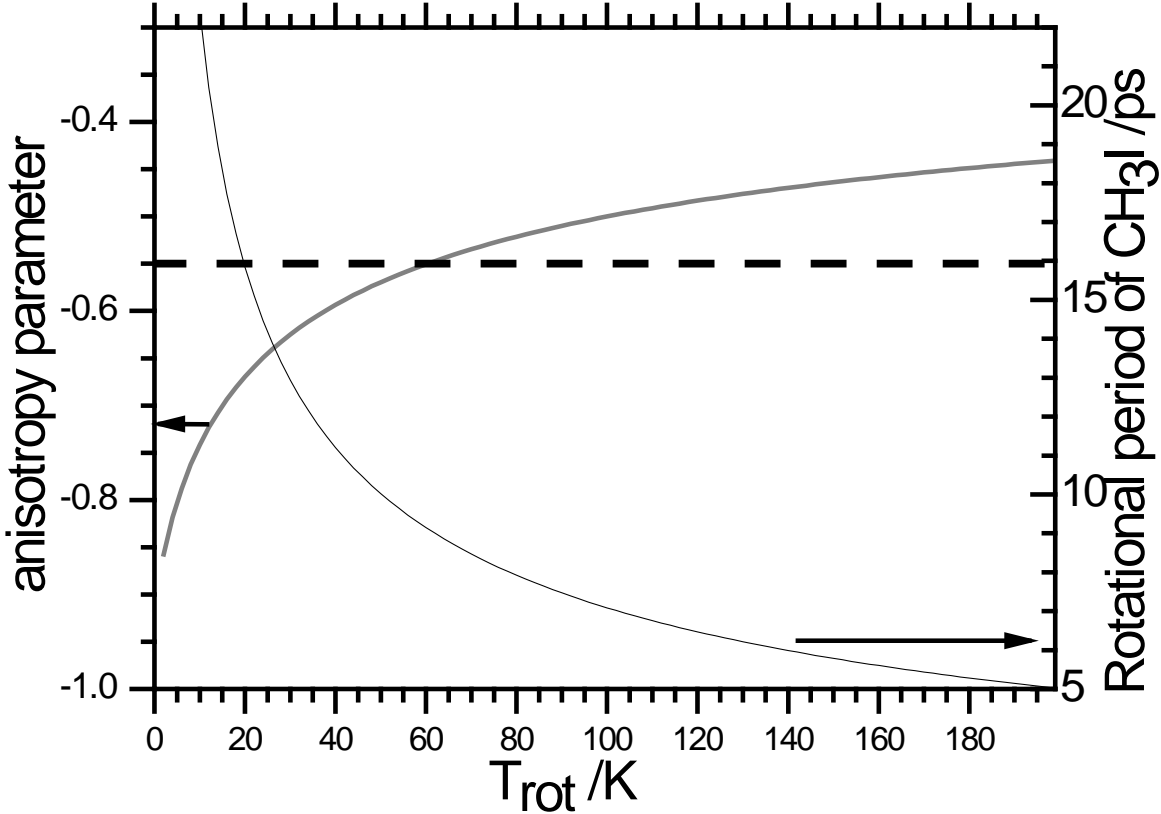